\documentclass[9pt,conference]{IEEEtran}
\usepackage{dcase2026}
\usepackage{bm} 
\usepackage{amsmath,graphicx,xurl,times,booktabs, tabularx,comment, subcaption}
\usepackage[table]{xcolor}

\usepackage{hhline}
\usepackage{hyperref}
\usepackage{xcolor}
\usepackage{cite}
\usepackage{multirow,url,hyperref}
\usepackage{siunitx}
\usepackage{cite}

\usepackage{pifont}
\usepackage{multirow}
\usepackage{comment}

\usepackage{hyperref}
\usepackage{xcolor}

\usepackage{etoolbox}
\makeatletter
\pretocmd{\thebibliography}{\footnotesize}{}{}
\makeatother

\title{Domain-Agnostic Incremental Learning for Sound Classification. \\ A DCASE 2026 Challenge task}

\name{Riccardo Casciotti$^{1}$,
       Manjunath Mulimani$^{2}$,
       Manu Harju$^{1}$,
       Jesper Rindom Jensen$^{2}$,
       Annamaria Mesaros$^{1}$
       }
 \address{$^1$ Signal Processing Research Centre, Tampere University, Finland \\
 $^2$ Department of Electronic Systems, Aalborg University, Denmark\\
 \{riccardo.casciotti, manu.harju, annamaria.mesaros\}@tuni.fi, 
 \{manjunathm, jrj\}@es.aau.dk\\
  }

\begin{document}

\maketitle

\begin{sloppy}

\begin{abstract}

This paper presents the \textit{ Domain-Agnostic Incremental Learning for Audio Classification} Task of the DCASE 2026 Challenge. Incremental learning refers to sequentially learning new tasks with the same system while maintaining its knowledge and performance on the previously learned task. Domain-incremental learning for sound classification refers to learning the same sound classes but in different acoustic domains, and was formalized as a data challenge for the first time in DCASE 2026. The task required participants to train a system to learn ten sound classes in three different domains, starting from a baseline system trained on the first domain. At each incremental task, learning had to be done without access to previous task data. The systems were ranked by the overall average accuracy calculated over the three domains. 
The task generated high interest and received submissions from 21 teams, with most submissions outperforming the provided baseline system. The highest performance of 79.6\%  was obtained using a combination of continual learning strategies that resulted in a balanced accuracy across domains, hence a good balance of retaining old and learning new information.

\end{abstract}

\begin{IEEEkeywords}
DCASE Challenge, Incremental learning, Sound event classification 
\end{IEEEkeywords}

\section{Introduction}
\label{sec:intro}

Incremental learning, also referred to as continual or lifelong learning in some contexts, studies how machine learning models can acquire new knowledge over time without forgetting what they previously learned. 
The \textit{Domain-agnostic incremental learning for Audio Classification} task was introduced in the DCASE Challenge 2026 for the first time. The task addresses a realistic and important scenario in machine listening, where audio classification systems are required to learn sounds from a sequence of successive domains, while maintaining performance on previously encountered domains.

Domain-incremental learning (DIL) is broadly categorized into two types: domain-aware and domain-agnostic incremental learning \cite{mulimani2025dil_icassp}. Domain-aware methods assume that domain identity is available during inference, allowing the model to select domain-specific parameters. Domain-agnostic incremental learning requires the model to classify the test audio samples from all domains seen so far without explicit domain identity, making it a more challenging and practically relevant setting.  
An overview of the proposed domain-agnostic incremental learning (DAIL) is illustrated in Fig.~\ref{fig:overview}.

Domain incremental learning differs from domain adaptation methods, which are restricted to only two domains: source and target \cite{gharib2018unsupervised}. Domain adaptation transfers knowledge from the source to the target domain and only considers the performance of the target domain after adaptation. Domain adaptation holds the data of the source domain to compare its distribution with the target domain. In comparison with domain adaptation, the domain incremental learning setup has multiple domains to adapt over time without accessing the previously seen domains. It takes the required measures to mitigate forgetting and considers the performance over all the seen domains so far. 

Incremental learning has been studied in the broader continual learning literature in different domains such as image classification \cite{kirkpatrick2017ewc,rebuffi2017icarl}, natural language processing \cite{dautume2019episodic}, and audio \cite{mulimani2023taslp, Xiao2022}. Mitigation of the catastrophic forgetting phenomenon, i.e. forgetting previously learned information when learning a new task, is implemented with approaches like regularization-based methods \cite{kirkpatrick2017ewc, li2017learning}, replay-based strategies \cite{rebuffi2017icarl, Xiao2022, wang2019continual}, and dynamic architectures \cite{li2025addressing}.

\begin{figure}[t!]
\centering
\includegraphics[width=.85\linewidth]{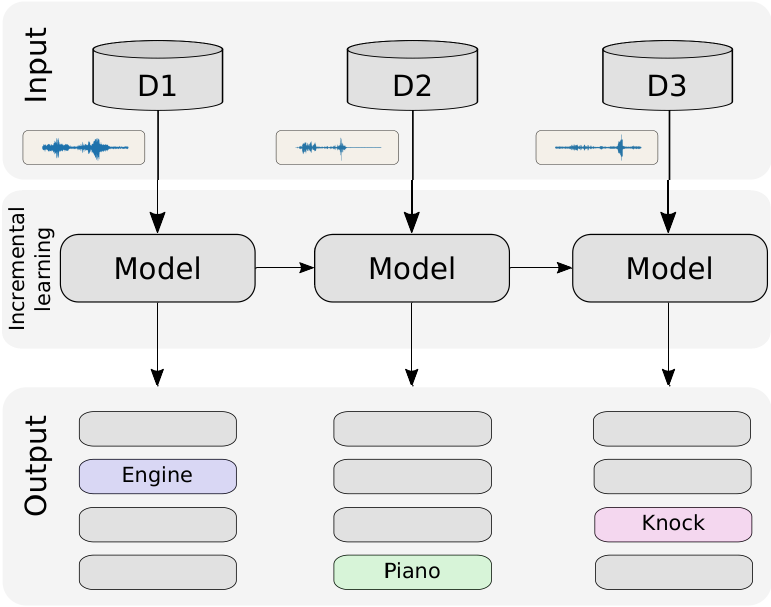}

\caption{Overview of \textit{Domain-agnostic incremental learning}. A model is trained incrementally with data consisting of a given set of classes but from three different domains. At each incremental stage, the model provided as baseline has access only to the current domain data. At inference time, the model has no domain knowledge and must classify the audio into one of the predefined target sound classes, irrespective of its domain.}
\label{fig:overview}
\end{figure}

In the audio domain, recent work has demonstrated the effectiveness of incremental learning for acoustic scene classification \cite{mulimani2021location}, sound event detection and localization \cite{pandey2025selddil}, and audio tagging \cite{mulimani2023taslp} with scenarios that cover class-incremental (learning new classes in successive tasks) and domain-incremental learning (learning the same classes from new domains) \cite{van2019three}.  
Domain incremental learning was also used as a learning procedure for obtaining generalization across locations \cite{mulimani2021location} and devices \cite{mulimani2024devices} for acoustic scene classification, using domain-specific normalization, low-rank adaptation, and knowledge distillation. 

Building on these studies, we formalized a benchmark task for domain-agnostic incremental learning for sound classification. The task setup requires systems to learn audio data from multiple domains sequentially, without access to data from previous domains and without domain labels at the inference stage. This setting discourages fine-tuning and encourages research into learning strategies that promote long-term knowledge retention and generalization. By introducing this topic as a dedicated challenge task, DCASE provided for the first time a common evaluation framework and dataset for systematic comparison of solutions. 

The remainder of the paper is organized as follows: 
Section~\ref{sec:setup} presents the task setup in detail, including the dataset, task rules, evaluation and baseline system. Section~\ref{sec:results} presents the baseline system results; after the challenge submission deadline, this section will be extended to present the challenge outcome, comparison of results and analysis of submitted solutions. Finally, Section~\ref{sec:conclusion} presents conclusions and future development ideas.

\section{Task Setup}
\label{sec:setup}
In this task, continual learning is performed through incremental exposure to sound events from different domains. Models must adapt to each new domain while maintaining stable performance on earlier ones, promoting long‑term learning and generalization.
Participants will train a sound event classifier under a domain‑incremental learning (DIL) scenario. Audio data from different domains D is presented sequentially, and at each stage the system must learn the new domain using only its data. 

\subsection{Dataset}

The task is constructed around three different domains, with the original datasets from which the sounds were selected including e.g. AudioSet \cite{gemmeke2017audio}. The data is presented as belonging to domains D1, D2 and D3, without reference to their original provenance. 
The data is annotated with ten classes:
\textit{alarm,
baby cry,
bark,
engine,
fire,
footsteps,
knock,
telephone ringing,
piano,} and
\textit{speech}.

The provided DIL-DCASE26 development dataset contains data from two different domains, with 139 minutes of audio from D2 and 275 minutes of audio from D3. The knowledge of D1 was provided in the baseline system: the system was provided readily trained on D1, and no access to audio from this domain is provided. 

Reference labels provided with the development data include the class label and domain. Each clip is annotated with only one label. Additional sounds may be present in the audio, but there is one target sound for the classification task. The development dataset was provided with a training/test split for reporting results at the development stage. 
The evaluation set contains files from all three domains, with only audio being provided, without domain information.

\subsection{Task rules, external data resources and pretrained models}

Use of external data and pretrained models that produce embeddings was not allowed. The intention was that the system development uses only data specifically provided in the task, for a fair comparison of the proposed approaches. 
However, manipulation of the provided development data was allowed, e.g. by mixing data sampled from a pdf or using augmentation techniques such as pitch shifting or time stretching.

Participants were not allowed to make subjective judgments of the evaluation data, nor to annotate it. Using the evaluation data to train the submitted system was not allowed, and the use of statistics about the evaluation data in the decision-making was also forbidden, such that the classification decision is done independently for each test sample.

\subsection{Baseline system}

The baseline system implements a convolutional neural network-based approach, in which learning of new domains is based on adjusting the batch normalization parameters (BN layers) to reflect the data distribution of each new domain. The baseline system is based on the work in \cite{mulimani2025dil_icassp}, and uses the domain-agnostic version of the system.

The baseline system architecture consists of 6 convolutional blocks, each including 2 convolutional layers followed by a batch normalization (BN) layer; the layer specifications are the same as PANNs CNN14 \cite{kong2020panns}. Global pooling is applied to the last convolutional layer, to obtain a fixed-length input feature vector that feeds into the classifier equipped with a softmax activation layer. 
The system was trained for 120 epochs using a batch size of 32, with data shuffling between epochs. The learning uses Adam optimizer, with a learning rate at the initial phase of 0.0001, and 0.00001 at incremental phases. 

For training the system, the audio clips were resampled to 32 kHz if necessary, then segmented into 4-second clips. 
Each clip was represented using log mel-band energies in 64 bands, with lower and upper cut-off frequencies 50 Hz and 14 kHz respectively, using a Hamming window of 1024 samples with 320 samples hop size in the analysis.
The baseline model was trained from scratch on D1, and provided with the knowledge embedded into it. For incremental learning of new domains D2 and D3, separate domain-specific BN layers were adapted for each domain. Figure~\ref{fig:batchnorm} illustrates this learning procedure.

\begin{figure}[t!]
\centering
\includegraphics[width=0.7\linewidth]{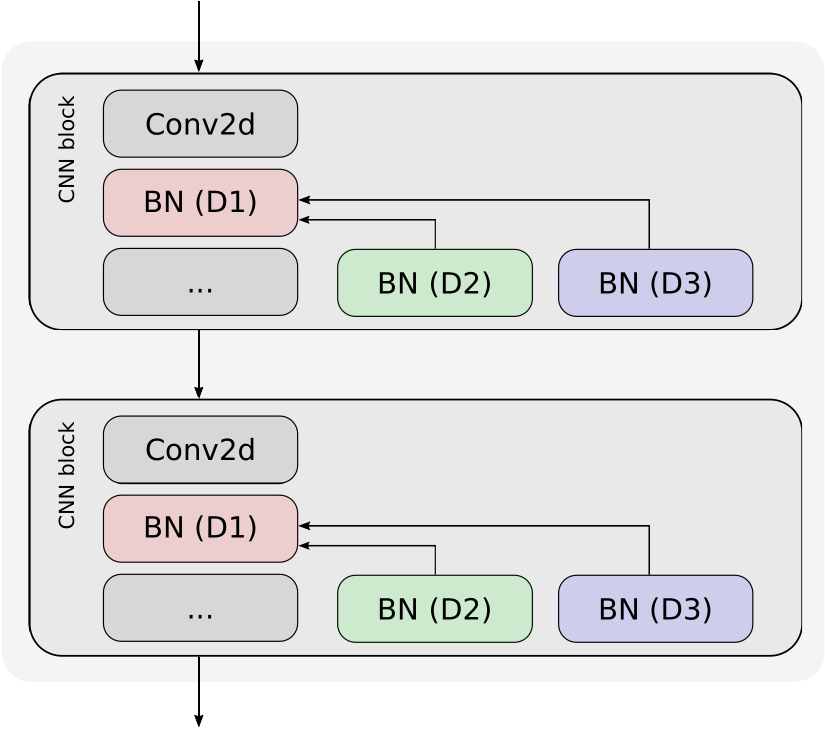}
\caption{Batch Normalization layers in the baseline model: for each domain D, a specific BN layer is trained using the data of the corresponding domain.}
\label{fig:batchnorm}
\end{figure}
During inference, domain-specific BN layers were predicted and used with domain-shared layers for classification. Specifically, an input audio was forward-passed through a combination of shared and domain-specific layers of each domain seen so far, obtaining the class probabilities. Subsequently, the uncertainty of the model on the given input audio among the predicted probabilities was computed using entropy. The domain-specific layers that have minimum entropy, denoting lower uncertainty, were selected for the final classification decision.

\subsection{Baseline system results}
During the development phase, the baseline system results were provided only on D2 and D3. After learning D2, the baseline has an accuracy of 58.6\% over the 8 classes in the test data of D2. After incrementally learning D3, the system has an accuracy of 46.1\% on the test data of D3 (9 classes), and shows a very slight increase in performance on D2 to 59.0\%. The average accuracy of the baseline model over D2 and D3 is 52.5\%. 

With a task-dependent approach, where the correct BN layers are selected for each test item, the performance is 70.6\% on D2 after learning it; after learning D3, performance on D3 is 56.9\%, and performance on D2 is maintained; the average across the two domains is 63.8\%. 
This indicates that more advanced methods for accurate prediction of domain-specific layers should significantly improve the overall performance. 

\section{Challenge results and analysis}
\label{sec:results}

The challenge received a high number of submissions, the total number of participating teams was 21. Each team was allowed to submit a maximum of 4 systems, but not all did so; the total number of submitted systems is 59.

\subsection{Evaluation and ranking}

Systems were ranked by \textit{overall accuracy} over the three domains present in the evaluation set. 
Accuracy was calculated separately for each domain, then overall accuracy was calculated as the average over the three domains. This way, the three domains have equal weight in the final ranking, accounting for potential data imbalance. Domain-wise accuracy was calculated as the average of the class-wise accuracies.

\subsection{Continual learning solutions}

We analyze the methods submitted to the task by first classifying them based on the continual learning solutions adopted by the participants. We can mainly distinguish four continual learning approaches that are widely adopted in the literature: regularization, replay, optimization, and architectural.
Regularization uses additional constraints to balance the old and new tasks, often using the frozen model information to preserve past knowledge. Replay approximates previous data distributions typically by keeping and using some samples of the old tasks in learning; however, only generative replay was allowed based on the task rules, where an external generative model provides samples generated from previously seen tasks. Optimization modifies the training procedure itself to reduce forgetting, while architectural methods introduce task-specific parameters to limit inter-task interference, either through parameter isolation or dynamic architectures.

The systems submitted to the challenge use individual or combinations of these solutions: 9 out of all the 21 teams submitted systems that use one continual learning solution alone, and 12 out of 21 adopt a hybrid approach combining different solutions. Specifically, 20 out of 21 teams used an architectural approach, 8 out of 21 used regularization, 5 out of 21 used optimization, and one team used a generative replay solution.

\begin{table}[t]
\centering
\caption{DCASE 2026 Teams ranking based on the average accuracy [\%] over the three domains.  Individual domain accuracies presented for analysis. Values marked with * are outside of the 95\% confidence interval of the next ranked system.}
\vspace{8pt}
\label{tab:dcase2026_task7_teams_2}
\begin{tabular}{cl|l|ccc}
\toprule

\textbf{Rank} & \textbf{Team} & \textbf{Avg. Acc} & \textbf{D1} & \textbf{D2} & \textbf{D3} \\

\midrule
1  & Park\_GIST-HanwhaVision & 79.6* & 87.1 & 81.5 & 70.3 \\
2  & Chun\_Chosun        & 78.8 & 91.7 & 79.7 & 62.4 \\
3  & Im\_SSU     & 76.9 & 88.8 & 78.6 & 63.3 \\
4  & Chang\_Surrey        & 75.1* & 80.3 & 82.6 & 62.5 \\
5  & Divakar\_IND    & 72.8 & 89.2 & 76.3 & 52.8 \\
6  & Miyazaki\_CyberAgent       & 72.3 & 88.0 & 70.0 & 59.0 \\
7  & Chang\_HYU             & 71.3 & 90.2 & 68.4 & 55.1 \\
8  & BelHadj\_VUB       & 70.7 & 85.7 & 67.9 & 58.3 \\
9  & Guan\_HEU      & 69.9 & 79.3 & 70.6 & 59.9 \\
10 & Li\_SCUT       & 69.4 & 86.8 & 68.1 & 53.3 \\
11 & Zhang\_XJTLU            & 69.1 & 81.5 & 69.3 & 56.4 \\
12 & Gao\_SHNU         & 68.9 & 87.1 & 59.5 & 60.0 \\
13 & Huang\_WHU        & 68.8 & 73.8 & 72.2 & 60.3 \\
14 & Heo\_SeoulTech         & 66.7 & 72.7 & 82.0 & 45.3 \\
15 & Chung\_IND         & 65.5 & 89.1 & 60.8 & 46.5 \\
16 & Tyagi\_IITM           & 63.1 & 81.5 & 67.0 & 40.8 \\
17 & Takami\_OU           & 62.4 & 86.1 & 53.3 & 47.8 \\
18 & Giacomini\_ICMC        & 62.3 & 82.9 & 56.5 & 47.7 \\
19 & Yang\_USST      & 62.3 & 86.4 & 55.1 & 45.5 \\
20 & Hu\_XJTLU        & 61.6 & 72.8 & 60.2 & 51.7 \\
 \cellcolor{gray!50}& \cellcolor{gray!50}Baseline System & \cellcolor{gray!50}60.3 & \cellcolor{gray!50}89.2 & \cellcolor{gray!50}49.7 & \cellcolor{gray!50}42.2 \\
21 & Raja\_IITM        & 59.7 & 77.6 & 41.1 & 60.2 \\
\bottomrule
\vspace{8pt}
\end{tabular}
\end{table}

\subsection{Architectural variants}

Most DCASE 2026 Task 7 submitted systems are built around the official CNN14/MCnn14 baseline architecture, or around lightweight modifications of it, typically adding domain-specific batch normalization \cite{Choi2026, Bel-Hadj2026, Raja2026a}, adapters or LoRA modules \cite{Bel-Hadj2026, Divakar2026, Miyazaki2026}, expert branches \cite{Park2026}, prototype classifiers \cite{Park2026}, routing mechanisms \cite{Choi2026, Miyazaki2026}, or ensemble strategies \cite{Park2026, Divakar2026}. However, some submissions introduce additional non-baseline components such as transformer-based fusion \cite{Tyagi2026}, multi-feature cross-attention modules \cite{Giacomini2026}, ridge/prototype readouts \cite{Pi2026, Raja2026a}, or explicit CRNN-LoRA variants \cite{Pi2026}. Therefore, a safer characterization is that the challenge is largely CNN14/MCnn14-baseline-centered, with a minority of systems extending it through CRNN \cite{Pi2026}, transformer \cite{Tyagi2026}, prototype \cite{Park2026, Pi2026}, or feature-fusion variants \cite{Giacomini2026}.

\begin{figure*}[t!] 

   \includegraphics[width=\textwidth]{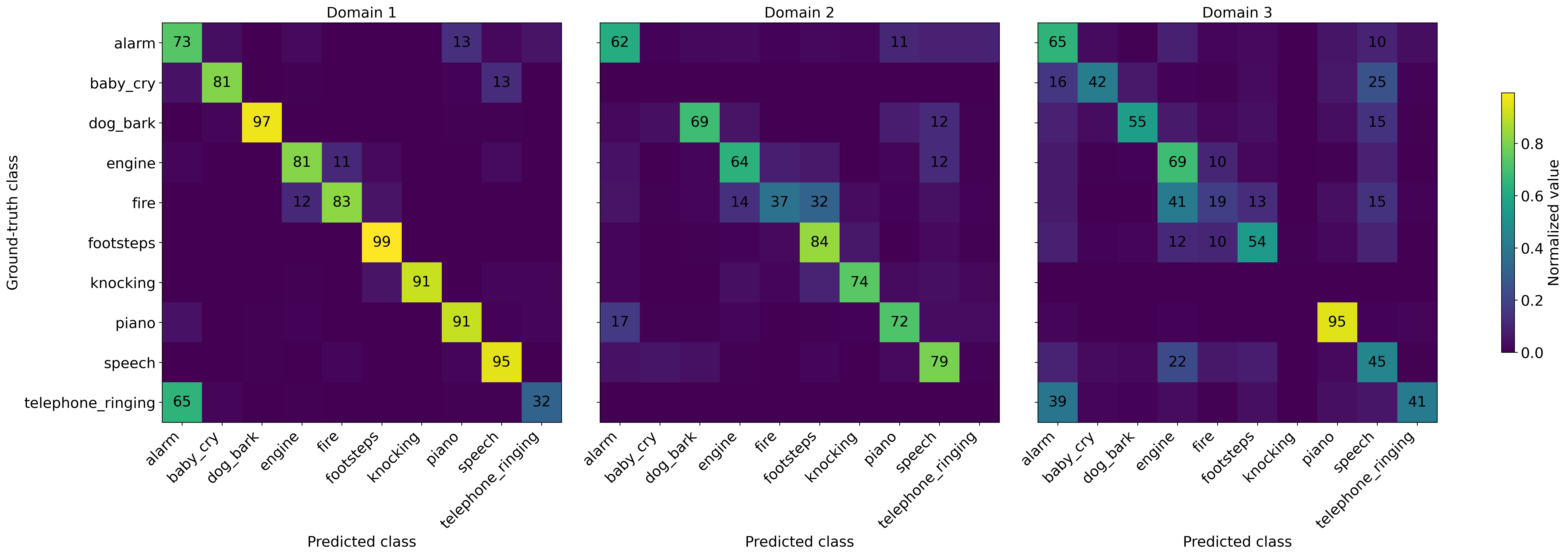}
   \vspace{2pt}
   \caption{Confusion matrices for each domain, calculated by aggregating the predictions of all submitted systems and then normalized. Each normalized row represents the distribution of predicted classes for a given ground-truth class, while the diagonal entries correspond to class-wise recall.}
   \label{fig:cf}

\end{figure*}

\subsection{Data augmentation strategies}

Data augmentation was used unevenly across the 
submissions. Many systems relied mainly on the official baseline preprocessing pipeline \cite{Chung2026} and focused on architectural or inference-time modifications rather than extensive augmentation \cite{Tyagi2026, Takami2026, Raja2026a}. When augmentation was used, it was generally designed to improve robustness to domain shift: examples include device simulation or device-related augmentation for Domain 2 \cite{Heo2026}, gain-shift augmentation for Domain 3 \cite{Heo2026}, SpecAugment \cite{Hu2026, Yang2026}, MixUp or Soft Mixup \cite{Hu2026, Giacomini2026}, frequency-statistics mixing such as Freq-MixStyle \cite{Hu2026}, and test-time augmentation with prediction averaging \cite{Heo2026, Yang2026}. Overall, the augmentation strategies were mostly conventional audio augmentations adapted to continual and domain-incremental learning, with the main goal of improving cross-domain robustness while limiting forgetting of previous domains.

\subsection{Overall results}
The evaluation results are shown in Table \ref{tab:dcase2026_task7_teams_2}, with only the best submission per team being shown. The complete list of systems is available on the task page\footnote{https://dcase.community/challenge2026/task-domain-agnostic-incremental-learning-for-audio-classification-results}. The table also includes the accuracy for each individual domain. 
Overall, the majority of the submissions are above the provided baseline, which sits second from the bottom in the teams ranking. 

From the results in Table \ref{tab:dcase2026_task7_teams_2} we observe that the top-5 performing teams have a more stable accuracy across domains; a higher variation in performance between domains 1 and 2 is seen in ranks 2 and 5, of over 10\%; the gap to domain 3 is typically much larger.
This suggests that the trade-off between memory-retention and learning plasticity has reached a good enough equilibrium in the top systems. The teams on mid to lower ranks have a significant gap between accuracies across domains, suggesting excessive memory-retention or excessive learning plasticity.

Confidence intervals show that overall performance in terms of the average accuracy across domains degrades very smoothly with the official rank, with only a few systems showing statistically significant differences from the next in rank. Domain-wise confidence intervals were calculated using the binomial distribution, and the aggregate confidence intervals were calculated using the normal approximation. \\ 
The confusion matrices for all three domains are shown in Fig.~\ref{fig:cf}. Domains 1 and 2 show a strong diagonal behavior, with some confusion; in domain 1, \textit{telephone ringing} is wrongly classified as \textit{alarm} on nearly 40\% of the samples. In domain 2,  \textit{footsteps} is misclassified as \textit{fire} on 32\% of the samples, and \textit{piano} as \textit{alarm} on 17\% of the samples. 

In domain 3 we observe more confusion. The submitted systems also suggest domain 3 poses a serious challenge, with many systems having an accuracy around 50\% on it, with a significant gap to other domains. This hints that domain 3 has a much larger distribution shift compared to domains 1 and 2, and that many systems failed to correctly capture this shift, including the baseline. In Figure \ref{fig:cf}, we can see that in general the confusion is focused in a grid of 4 classes: \textit{fire}, \textit{telephone ringing}, \textit{speech}, and \textit{baby cry}. Some systems focused too much on retention in domains 1 and 2, while not giving enough plasticity to the model to successfully learn domain 3. Additionally, many systems such as \cite{Hu2026} focus on using domain-specific components, which require a strong routing solution in order to infer the correct domain, a bottleneck that negatively affected many architectural continual learning solutions submitted to the challenge.

\subsection{Analysis of the top-performing systems}

The top-performing systems 
indicate that strong performance requires more than standard continual-learning regularization. The best submissions combine three recurring ideas: preserving previous-domain knowledge, adding domain-specific adaptation capacity, and improving domain-agnostic routing at inference time. The first ranked team, Park\_GIST-HanwhaVision,  follows an expert-preservation strategy: compact experts are trained incrementally, frozen after each stage, and later combined through their penultimate representations using a lightweight prototype classifier \cite{Park2026}. DeepInversion-based replay \cite{yin2020dreaming, Park2026} and cross-stage feature imputation \cite{Park2026} further help compensate for missing expert features, reducing forgetting without requiring direct access to previous-domain data and obtaining a very balanced performance across the three domains.

The second-ranked team, Chun\_Chosun, follows a parameter-efficient alternative, freezing the CNN14 backbone and adapting to new domains through domain-specific low-rank adapters \cite{Chanjun2026}. Knowledge distillation and orthogonal gradient projection \cite{Chanjun2026} are used to constrain updates and reduce interference with previously learned domains.

The next best systems emphasize that domain-agnostic inference is itself a central difficulty. Im\_SSU uses a CNN14-based framework with calibrated soft routing, combining prototype distances, confidence, margin, entropy, and out-of-distribution evidence to estimate domain responsibility \cite{Choi2026}. Chang\_Surrey similarly starts from a frozen CNN14 model, but isolates new-domain learning through domain-specific batch normalization, low-rank convolutional adapters, and classifier deltas \cite{Chang2026}. Its final prediction is produced by router-weighted fusion of per-domain heads \cite{Chang2026}, combining architectural isolation with learned model selection. Ranked 5th, \cite{Divakar2026} addresses the limited adaptation capacity of the baseline by inserting residual bottleneck adapters after CNN14 convolutional blocks. Their system assigns deeper adapters to the more challenging Domain 3 and uses a supervised contrastive auxiliary loss to improve feature separability.

Overall, the top-five approaches show that the most effective systems avoid directly fine-tuning a single shared model across all domains. Instead, they freeze the backbone, preserve domain experts, isolate new parameters, or constrain updates to limit interference. At the same time, they rely on routing, prototype matching, or head fusion to solve the domain-agnostic inference problem. The persistent drop on domain 3 indicates that future improvements may depend less on generic anti-forgetting regularization and more on reliable domain-confidence estimation, adaptive expert fusion, and domain-specific capacity allocation.

\section{Conclusion}
\label{sec:conclusion}

This paper presented DCASE 2026 Task 7, the first DCASE Challenge task on continual learning, formulated as domain-agnostic incremental sound classification. Systems had to learn the same classes across sequential domains without previous-domain data or domain labels at inference. The results show that the task is feasible but challenging: most submissions outperformed the baseline, especially those using architectural strategies. However, domain-agnostic routing remains a major bottleneck, with domain 3 causing the largest performance drop and several classes showing strong confusion.  
Overall, the task provides a benchmark for robust incremental audio classification and highlights the need for improved retention–plasticity trade-offs, domain-confidence estimation, and adaptive expert fusion.

 \newpage
\bibliographystyle{IEEEtran}
\bibliography{refs}

\end{sloppy}
\end{document}